%% file: auctions.tex
\documentclass[conference]{IEEEtran}

\usepackage[utf8]{inputenc}
\usepackage{hyperref}
\usepackage{listings}
\usepackage{subcaption}
\usepackage{soul}
\usepackage{times}
\usepackage{array,multirow}
\usepackage{xspace}
\usepackage{xcolor}
\usepackage{color, colortbl}
\usepackage{balance}
\usepackage{paralist}
\usepackage{amsfonts,amssymb}
\usepackage{amsthm,amsmath}
\usepackage{cleveref}
\usepackage{graphicx}
\usepackage{pifont}
\usepackage{array}
\usepackage{booktabs}
\usepackage{tikz}
\usepackage{bm}
\usepackage{todonotes}

\let\labelindent\relax
\usepackage{enumitem}

\usepackage[font=footnotesize,labelfont=bf]{caption}
\input{macros.tex}

\renewcommand{\baselinestretch}{0.98}

\begin{document}
%
\title{\rname: Auction-based Shared Economy ResolutIon System for blocKchain}

\author{
\IEEEauthorblockN{Alberto Sonnino}
\IEEEauthorblockA{University College London\\alberto.sonnino@ucl.ac.uk}
\and
\IEEEauthorblockN{Michał Król}
\IEEEauthorblockA{University College London\\m.krol@ucl.ac.uk}
\and
\IEEEauthorblockN{Argyrios G. Tasiopoulos}
\IEEEauthorblockA{University College London\\argyrios.tasiopoulos@ucl.ac.uk}
\and
\IEEEauthorblockN{Ioannis Psaras}
\IEEEauthorblockA{University College London\\i.psaras@ucl.ac.uk}
}

\maketitle

\begin{abstract}
Recent developments in blockchains and edge computing allows to deploy decentralized shared economy with utility tokens, where altcoins secure and reward useful work. However, the majority of the systems being developed, does not provide mechanisms to pair workers and clients, or rely on manual and insecure resolution. \rname bridges this gap allowing to perform sealed-bid auctions on blockchains, automatically determine the most optimal price for services, and assign clients to the most suitable workers. \rname allows workers to specify a minimal price for their work, and hide submitted bids as well the identity of the bidders without relying on any centralized party at any point. We provide a smart contract implementation of \rname and show how to deploy it within the Filecoin network, and perform an initial benchmark on \chainspace.
\end{abstract}

\input{sections/introduction.tex}
\input{sections/background.tex}
\input{sections/goals.tex}
\input{sections/design.tex}
\input{sections/security.tex}
\input{sections/evaluation.tex}
\input{sections/related.tex}
\input{sections/limitations.tex}
\input{sections/conclusion.tex}

\input{sections/ack.tex}

\bibliography{biblio} 
\bibliographystyle{ieeetr}

\end{document}

%% file: macros.tex
\newcommand\alberto[1]{\todo[inline,color=yellow]{\textbf{Alberto:} #1}}

\newcommand{\rname}{AStERISK\xspace}
\newcommand\sysname{AStERISK\xspace}

\newcommand\chainspace{Chainspace\xspace}

\newcommand\coconut{Coconut\xspace}

\newcommand\algorithm[1]{\textsf{#1}}

\newcommand\function[1]{\ding{118}\xspace \textsf{#1}:\xspace}

\newcommand\bidder{bidder\xspace}
\newcommand\bidders{bidders\xspace}

\newcommand\Bidders{Bidders\xspace}
\newcommand\authority{authority\xspace}
\newcommand\authorities{authorities\xspace}

\newcommand\worker{worker\xspace}
\newcommand\workers{workers\xspace}

\newcommand\Workers{Workers\xspace}

\newcommand{\etal}{\textit{et al.}\@\xspace}
\newcommand{\eg}{\textit{e.g.,}\@\xspace}
\newcommand{\ie}{\textit{i.e.,}\@\xspace}

\newcommand{\cmark}{\ding{51}}%
\newcommand{\xmark}{\ding{55}}

\def\first{({\it i})\xspace}
\def\second{({\it ii})\xspace}

\definecolor{verylightgray}{gray}{0.9}

\newcolumntype{L}{l<{\hspace{0.8cm}}}
\newcolumntype{C}{c}

\DeclareRobustCommand\pie[1]{
\tikz[every node/.style={inner sep=0,outer sep=0, scale=1.5}]{
\node[minimum size=1.5ex] at (0,-1.5ex) {}; 
 \draw[fill=white] (0,-1.5ex) circle (0.75ex); \draw[fill=black] (0.75ex,-1.5ex) arc (0:#1:0.75ex); 
}
}
\def\L{\pie{0}} 
\def\M{\pie{-180}} 
\def\H{\pie{360}} 

%% file: sections/introduction.tex
\section{Introduction}
The cloud computing model developed during the last two decades was built on the premise of compute centralization. That is, computing power is geographically and administratively concentrated in compute infrastructures of industrial scale, generally called datacenters. As a result, the majority of users currently rely on clouds for applications such as hosting services, offloading computation, and data storage. However, centralization introduces several drawbacks; cloud computing services act as large central points of failure~\cite{awsdown}, and make possible for authorities to enforce censorship~\cite{awscensor} or violate user privacy~\cite{debatin2009facebook}. Furthermore, large cloud operators often abuse their market position to effectively force users to trust their service and adapt to operators' rules and prices. 

Recent research efforts focus on shared-economy infrastructures, where services are performed by users sharing their resources with each other at the edge of the network \cite{abbas2018mobile}. Such infrastructures can provide better quality of service, and reduce the exposure to central points of failures and abuse of power; but require security solutions and reliable incentive mechanisms to compensate the lack of trust between distributed users. Following the recent success of Bitcoin~\cite{bitcoin} a large amount of cryptocurrency projects implement such a shared economy model and use blockchains to secure their platforms and simplify payments~\cite{golem, sonm, iexec, bounty0x, combinatorjune,protocol2017por}. The vision is to create a decentralized system, where users are incentivized to perform useful work and automatically receive rewards upon tasks completion. 

While multiple projects focus on the crucial task of proving to the network that a service has been successfully completed \cite{airtnt, krol2018spoc, miller2014permacoin, benet2018proof}, an equally important task of determining an optimal price for those services has been largely ignored by the community. Furthermore, industrial projects either do not cover this problem or rely on manual user interaction causing multiple scalability and security issues~\cite{golem, sonm, iexec, bounty0x}. In this work, we answer to the question  of \textit{how blockchains and smart contracts can be leveraged for deriving the price of a service in distributed computing infrastructures and automatically assign service requesters to corresponding workers.} 

Clearly, the providers of the distributed infrastructure are expected to ask for compensation, since admitting user requests will impose operational expenses while occupying their personal computing as well as storage resources. Therefore, the sufficient participation of infrastructure providers is associated with an incentivisation mechanism that allows them to profit by offering their resources at a price that equals or exceeds, their expenses. That is, the objectives of users and providers are conflicting with each other since the former try to access a service at the lowest possible price while the latter try to maximise their revenue. 
Therefore, there is a need for a market mechanism that will intervene, in the form of an \textit{auction}, to ensure the effective association between user requests to providers' resources. By running auctions on blockchain, user do not have trust each other nor any trusted 3$^{rd}$ party, they inherit security guarantees from the underlying distributed ledger and can be sure that auction is performed correctly using \emph{Smart Contracts}. However, data submitted to blockchain automatically becomes public and in sealed-bid auctions it is critical to hide all the bids during the auction. Moreover, revealing bidders identity imposes another security threat; \ie revealing that a bidder bought storage space at two servers makes the servers an easy target for malicious users willing the bidder to lose their data. 

In this paper we propose \rname --- an auction-based shared economy resolution system running on top of distributed ledger. In \rname, workers submit their offers to the blockchain together with a minimal price they are willing to accept. Our system hides submitted bids and protects bidders identity using anonymous credentials. In contrast to related work, \rname does not rely on a single trusted 3$^{rd}$ party to issue all the credentials, but rather on a set of multiple, decentralised authorities. Each user can define their own set trusted parties and is protected from a subset of authorities becoming malicious. Once an auction is finished, \rname automatically creates a binding on the blockchain allowing workers to claim money from requester deposits upon submission of a valid proof of useful work. For simplicity, we focus on the case of Filecoin~\cite{filecoin}, a decentralized storage network, but \rname can be easily adapted to work with additional systems implementing a shared economy model.

%% file: sections/background.tex
\section{Background}\label{sec:background}


\paragraph{Blockchain} The blockchain technology~\cite{bitcoin} implements a distributed, append-only ledger in the form of connected blocks; once information is stored in the blockchain it cannot be removed or altered. Network participants use a consensus protocol to agree on the current state of the ledger, and the system maintains its properties as long as a subset of the participants is honest. Blockchains are used to record crypto-currencies (\eg, Bitcoin~\cite{bitcoin}, Ethereum~\cite{buterin2013ethereum}) transactions. A common extension consisting of scripting languages enables to include logic as part of the transaction and allows deployment of smart contracts---code submitted to the blockchain and executed by all network participants. \sysname leverages the high-integrity data structure provided by blockchains, and uses it for accountancy, auditability, and time reference (\eg time may be define as the block height of the main chain).

\paragraph{Filecoin} At the core of Filecoin lies a Storage Market that allows requesters to pay miners, to store data. Specifically, a \emph{Storage Market} acts as an exchange point where clients and miners can advertise their requests and offer resources. The \textit{Network} guarantees that the miners are rewarded by the clients when providing the service. Filecoin uses Proof-of-Replication~\cite{benet2018proof} a scheme that relies on zero-knowledge Succinct Non-interactive ARguments of Knowledge (zk-SNARKs)~\cite{gennaro2013quadratic, ben2013snarks} allowing a Prover to prove possession of data $\mathcal{D}$. To prevent Sybil attacks and allow proving possession of multiple copies of the same file, Proofs-of-Replication are generated over a version of data encrypted under a key $ek$ specified by the requester such that $\mathcal{R}^\mathcal{D}_{ek} = \textsf{Encode}(\mathcal{D}, ek)$. Filecoin requires the Prover to recursively and continuously generate those proofs with randomness $\textit{r}$ obtained from the most recent block in the blockchain and thus proving possession of the file over time. 

\paragraph{Anonymous Credentials} Anonymous credentials~\cite{amac, pointcheval} allow the issuance of credentials to users, and the subsequent unlinkable revelation to a verifier. Users can selectively disclose some of the attributes embedded in the credential or specific functions of these attributes. Most anonymous credentials scheme entrust a single authority with a master credential signature key, allowing a malicious authority to forge any credential. Other schemes do not provide the necessary re-randomization or blind issuing properties necessary to implement general purpose disclosure credentials. To overcome these limitations, \rname relies on \coconut~\cite{coconut} which supports distributed threshold issuance of credentials; therefore supporting private attributes, re-randomization, and unlinkable multi-show selective disclosure without relying on a central trusted 3$^{rd}$ party. \coconut is designed for use in the context of blockchains to ensure confidentiality, authenticity and availability even when a subset of credential issuing authorities are malicious or offline; and uses short and computationally efficient credentials that can easily be verified by a smart contract. Colluding authorities can forge \coconut credentials, but cannot break unlinkability and de-anonymize users. \coconut authorities issue credentials without communicating with each other, following a standard key distribution phase; as a result, a large number of authorities may be used to issue credentials without significantly affecting efficiency.

%% file: sections/goals.tex
\section{Design Goals}\label{sec:goals}

\rname associates worker resources to users via the execution of an auction mechanism. Specifically, we consider a system with the following actors:
\begin{itemize}
\item \textbf{Bidders} - users willing to access a services (\eg Filecoin users wishing to store some data in the network).
\item \textbf{Workers} - offer their services to the users (\eg Filecoin miners wishing to store users file for specific prices).
\item \textbf{Authorities} - distributed system responsible to issue credentials allowing users to participate in auctions.
\end{itemize} 

\sysname assumes that at least a threshold subset of the authorities are honest; all cryptographic operations rely (or are implied by) the XDH assumption~\cite{bls}; and relies on weak synchrony\footnote{Weak synchrony~\cite{bft} is required by many smart contract platforms~\cite{chainspace, omniledger}, and by distributed key generation protocols required by \coconut~\cite{coconut}.} for liveness (but not for safety). These assumptions are inherited from \coconut, and the underlying smart contract platform. Given this threat model, \rname achieves the following design goals:

\begin{itemize}

\item \textbf{Hidden Minimum Price:} \Workers can specify a minimum price for which they are willing to perform given actions, and are guaranteed that the winner of the auction bids at least that price. This minimum price is kept private from the \bidders until the end of the auction.

\item \textbf{Bidders Privacy:} \Bidders are unlinkable to to their bids. Only the identity of the winner is revealed to the \worker (at the end of the auction).

\item \textbf{Bids Privacy:} Bids are kept private until the end of the auction; \bidders submit their bids without knowledge of what other \bidders do.

\item \textbf{Bids Binding:} \Bidders cannot change their bids once they are committed.


\item \textbf{Public Auditability:} Anyone can verify the correct execution of any auction.

\item \textbf{Fairness:} \Bidders are financially penalized if they deviate from the protocol, but cannot be financially penalized if they follow it correctly. \Bidders cannot double-spend coins~\cite{karame2012double}, and no \authority can steal \bidder's funds. 

\item \textbf{Non-Interactivity:} \Bidders are not required to interact with each other.

\item \textbf{Censorship Resistance:} Anyone can act as \bidder or \worker; the system is resilient to censorship.

\item \textbf{Distributed Authority:} \rname never relies on a single trusted 3$^{rd}$ party.

\item \textbf{Auction's economic properties: } Involving \textit{i}) \textit{Efficiency}, in terms of assigning resources to the \bidders that value them the most in a computationally feasible way, \textit{ii}) \textit{Incentive Compatibility} (\textit{Truthfulness}), where \bidders and \workers benefit by revealing their true valuations, \textit{iii}) \textit{Individual Rationality}, where both \bidders and \workers are willing to participate, and \textit{iv}) \textit{Budget Balance}, where the payments submitted cover \workers' compensations.
\end{itemize}

%% file: sections/design.tex
\section{\rname Design}\label{sec:design}

\begin{figure*}
\centering
\begin{subfigure}[b]{0.3\textwidth}
\includegraphics[width=\textwidth]{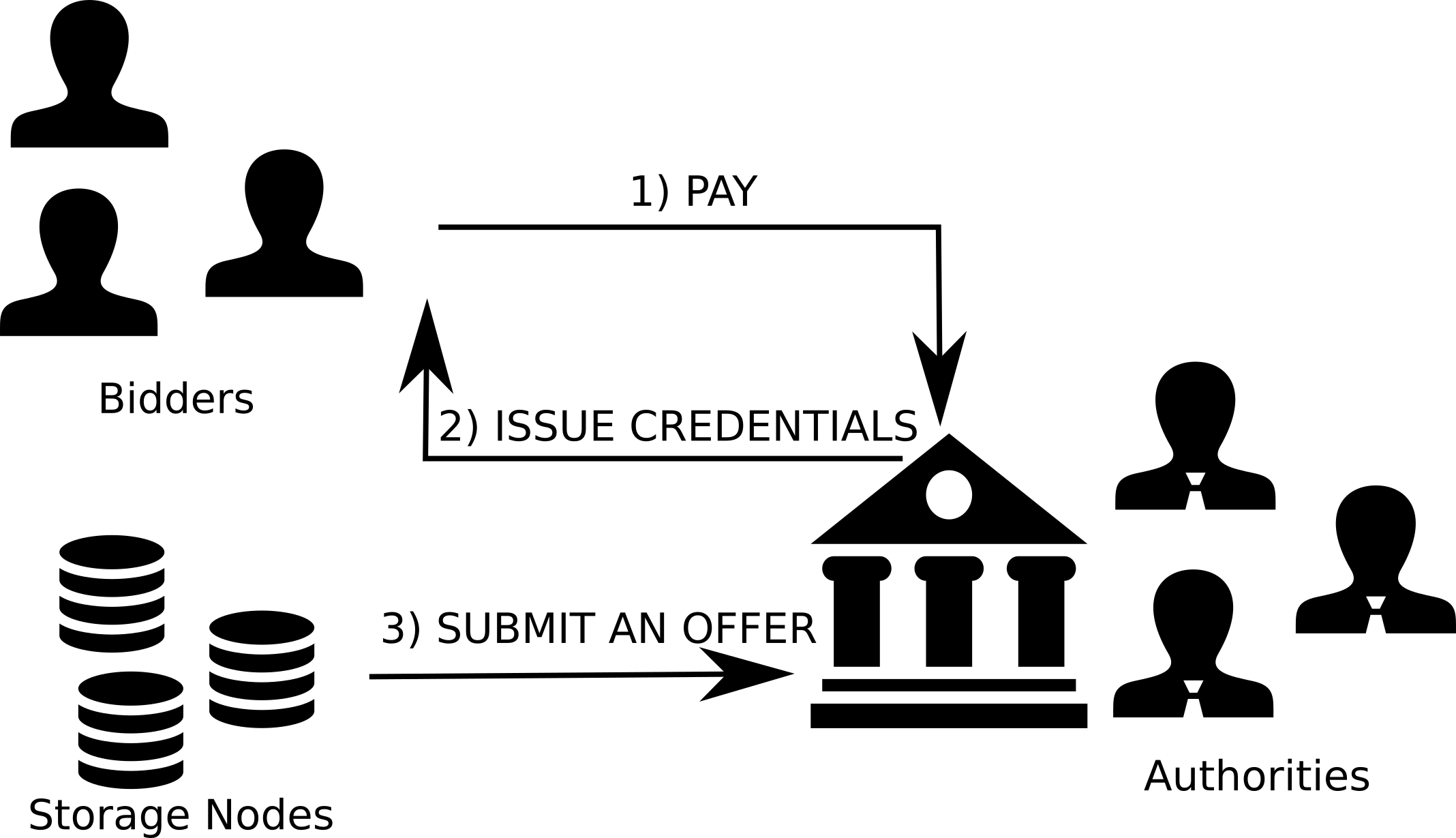}
\caption{Preparation phase.}
\label{fig:overview1}
\end{subfigure}
~ 
 \begin{subfigure}[b]{0.3\textwidth}
\includegraphics[width=\textwidth]{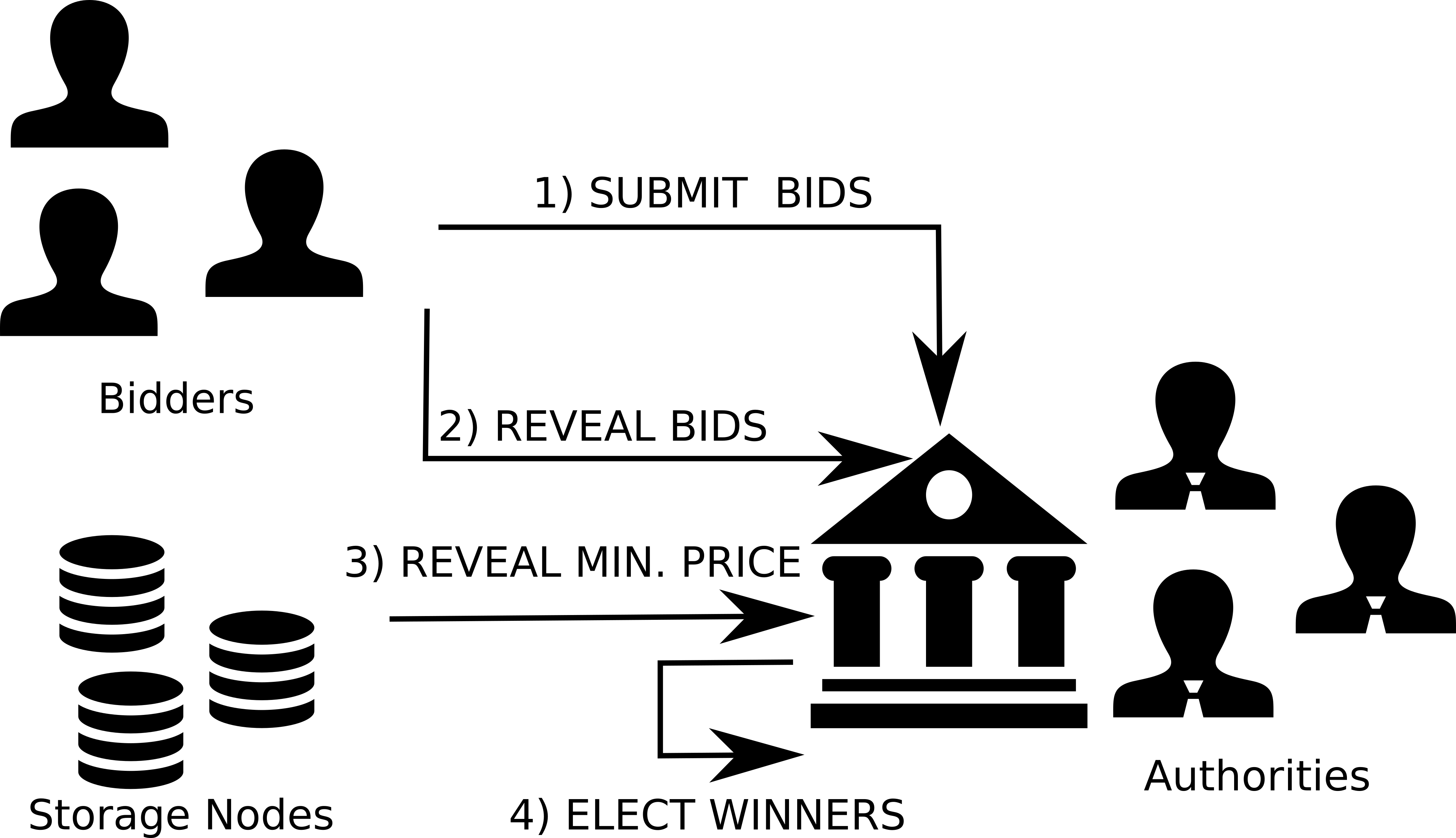}
\caption{Auction phase.}
\label{fig:overview2}
\end{subfigure}
~
\begin{subfigure}[b]{0.3\textwidth}
\includegraphics[width=\textwidth]{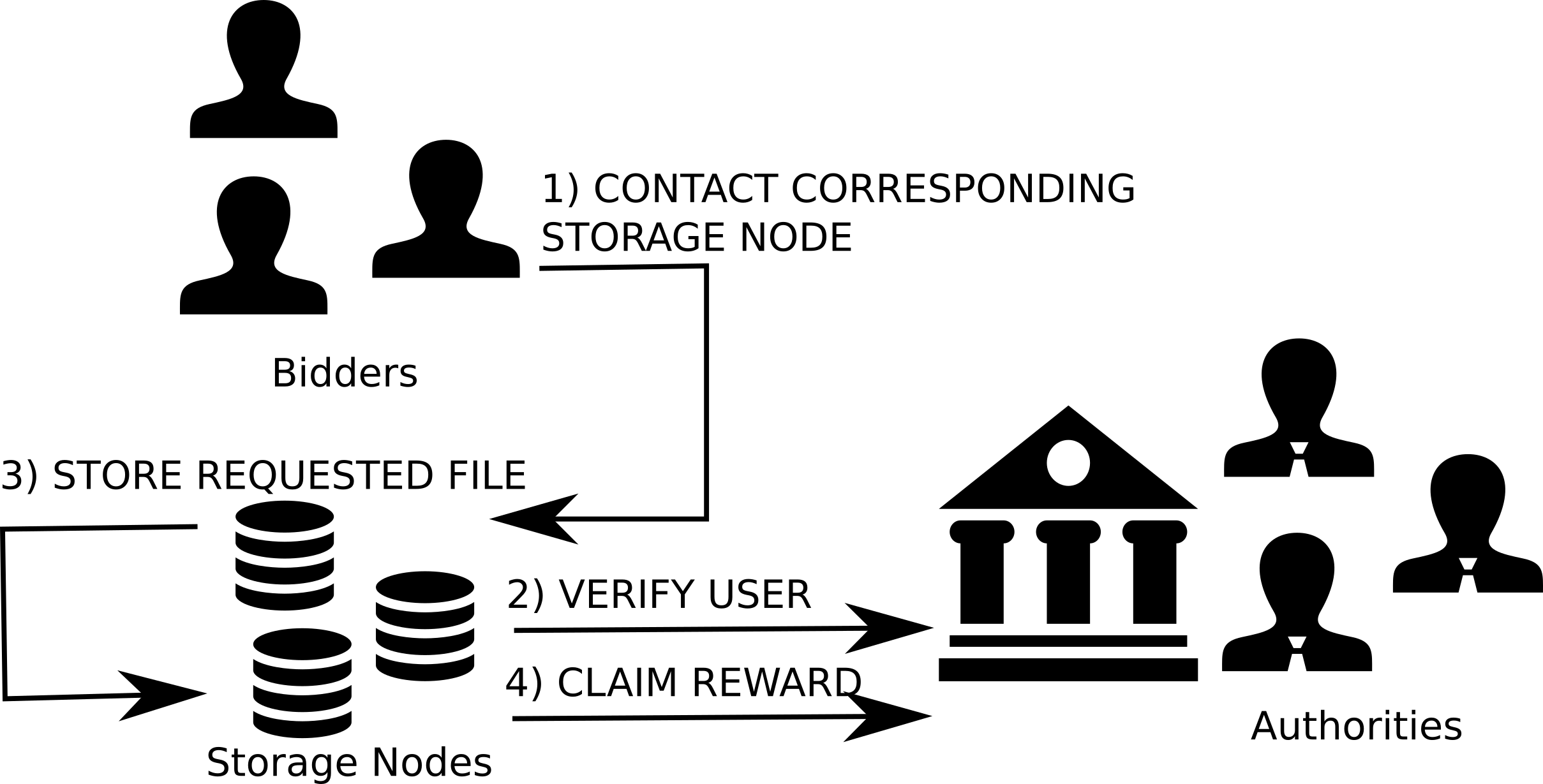}
\caption{Execution phase.}
\label{fig:overview3}
\end{subfigure}
\caption{Overview of auction executions on \rname}\label{fig:overview}
\vspace{-15pt}
\end{figure*}

\subsection{\rname Smart Contract} \label{sec:smart-contract}
We design \rname as a smart contract that extends the tumbler application of \coconut~\cite{coconut} to allow credentials to be used as anonymous bids in auctions\footnote{The tumbler application is described at Section V.A of \coconut~\cite{coconut}.}. The \rname smart contract defines six functions (\algorithm{Setup}, \algorithm{Create}, \algorithm{Deposit}, \algorithm{Commit}, \algorithm{Reveal}, \algorithm{Withdraw}):

\begin{description}[leftmargin=1em, labelindent=0em]
\setlength\itemsep{0.5em}

\item[\function{Setup}] A set of \authorities jointly create an instance of the \rname contract by providing their public keys as well as any other scheme parameters as the number of \authorities and the threshold parameter. This function can be run multiple time, and by different sets of \authorities; the \workers then select the set of authorities they trust upon executing \algorithm{Create}.

\item[\function{Create}] Any 3$^{rd}$ party \worker creates an auction by specifying the set of \authorities trusted to issue  credentials, as well as any application specific parameter or policy. They also specify, a commitment to the minimum price for which they are willing to operate; and two time-tamps\footnote{Time may be defined as the number of blocks built on top of the main chain of the blockchain.}, $t_{commit}$ and $t_{reveal}$ (where $t_{commit} < t_{reveal}$) used during the auction phase (see \Cref{sec:protocol}). 

\item[\function{Deposit}] Bidders deposit $v$ coins into a buffer account specified by the smart contract to request a credential on the public attribute $v$, and on a private randomly generated sequence number $s$. To prevent tracing traffic analysis, $v$ should be limited to a specific set of possible values, similar to cash denominations. Each authority monitors the \rname smart contract, and issue a partial credential to the user---either on chain or off-chain---upon detecting a credential request (credential requests are processed only if \bidders paid a deposit of $v$ coins to the smart contract). \bidders locally aggregate all partial credentials into a consolidated credential.


\item[\function{Commit}] \Bidders submit a bid by showing  a valid credential to the smart contract; they also provides a proof of knowledge of the sequence number $s$ along with a group element $\zeta$ uniquely built from $s$. If the proof and the credential check, the smart contract records $\zeta$. The group element $\zeta$ embeds the sequence number $s$ and is therefore bound to the credential and the number of coins $v$ it embeds---showing $\zeta$ effectively commits to $v$. The smart contract accepts the \bidders input only before $t_{commit}$.

\item[\function{Reveal}] \Bidders reveal $v$, $\zeta$, and the credentials, as well as a proof that $v$ is correctly embedded in the credentials and asserting correctness of $\zeta$. The smart contract accepts the \bidders input only after $t_{commit}$ and before $t_{reveal}$, and if the smart contract previously recorded $\zeta$ (\ie if the \bidder committed to its bid). \Workers open the commitment to the minimum price they committed to during \algorithm{Create}. 

\item[\function{Withdraw}] To withdraw the coins, the \bidder provides the smart contract with a zk-proof of knowledge of their private key by binding the proof to the address $addr$ where they wish to redeem the coins; they also provides the consolidated credential, $\zeta$, and a zk-proof attesting its correctness. To prevent double spending, the  contract keeps a record of all group elements $\zeta$ that have already been shown. Upon showing a $\zeta$ embedding a fresh (unspent) sequence number $s$, the contract verifies the credential and zero-knowledge proofs, and that $\zeta$ doesn’t already appear in the spent list. Then, it withdraws $v$ coins from the buffer and sends them to the specified address $addr$, and adds $\zeta$ to the spent list. \Bidders can only withdraw coins after $t_{reveal}$, and only credentials that have been recorded by \algorithm{Commit} and \algorithm{Reveal} can be used to withdraw coins (this effectively lock the funds of \bidders that deviate from the protocol). According to the policy set upon executing \algorithm{Create}, the winner of the auction may be treated differently.

\item[\function{SubmitWork}] After $t_{reveal}$, the winner contacts the \worker through a private channel,
and uploads cryptographic material to the smart contract allowing the worker to retrieve its payment. Winners can anonymously prove they actually won the auction by proving possession of the consolidated credential, $\zeta$, and a zk-proof attesting its correctness. In the case of Filecoin, the winner proves they won the auction to the smart contract, and uploads a hash of the encrypted file to store along with a signature. The file hash is then used to verify Proofs of Spacetime~\cite{benet2018proof} submitted by the worker and release payments from the \bidder's deposit over time.

\end{description}

\subsection{The \rname Protocol} \label{sec:protocol}
\Cref{fig:overview} shows the execution of an auction on \rname; the auction is divided in three phases:
\begin{itemize}
\item \textbf{Preparation phase (\Cref{fig:overview1}):} \Bidders pay deposits and retrieve a credential required to participate in auctions, and storage nodes submit their offers.
\item \textbf{Auction phase (\Cref{fig:overview2}):} \Bidders commit and later reveal their bids. The smart contracts determines the winner by applying a Vickrey auction mechanism~\cite{vickrey1961counterspeculation}.
\item \textbf{Execution phase (\Cref{fig:overview3}):} Auction winner contacts its corresponding storage nodes, prove their identity and submit files to store. Storage node receives rewards for storing files over time. 
\end{itemize}

\paragraph{Preparation phase}
A set of authorities executes the \algorithm{Setup} function of the \rname smart contract described in \Cref{sec:smart-contract}. Any \worker executes \algorithm{Create} (\Cref{fig:overview1}-\ding{204}) to create an auction by specifying the auction parameters, and which set of \authorities they trust. They also specify the two time-tamps, $t_{commit}$ and $t_{reveal}$ determining the auction's time line; \workers advertise their product and provide a commitment to a minimum price $v_0$ at which they are willing to operate. \Bidders execute \algorithm{Deposit} by paying a deposits of $v$ coins into a buffer account specified by the \rname smart contract (\Cref{fig:overview1}-\ding{202}), and retrieve a credential required to participate in auctions (\Cref{fig:overview1}-\ding{203}). The value $v$ represents the number of coins they wish to bid.

\paragraph{Auction phase}
\Bidders execute \algorithm{Commit} (\Cref{fig:overview2}-\ding{202}) to commit to a bid for a particular auction on the smart contract. \Bidders commitments are only considered valid if submitted before the deadline $t_{commit}$. Next, \workers open the commitment to their minimum price, and  \bidders call \algorithm{Reveal} (\Cref{fig:overview2}-\ding{203}) to open their commitment to the smart contract; \bidders openings are valid  only if submitted before the deadline $t_{reveal}$. Finally (\Cref{fig:overview2}-\ding{204}), the winner of the auction is deduced from the execution trace of the smart contract. The group element $\zeta$ associated with the highest $v$ (if $v\geq v_0$) indicates the winner of the auction---all \bidders that correctly followed the protocol (except the winner) may withdraw their coins calling the \algorithm{Withdraw} function of the \rname contract. If there is no winner, the auction fails and every \bidder may withdraw their money. \rname applies a Vickrey auction mechanism~\cite{vickrey1961counterspeculation}; the winner of the auction is the bidder with the highest bid $v$, and pays the price of the second highest bid, $v'$. That is, the winner is free to call \algorithm{Withdraw} to withdraw $(v-v')$ coins. 

\paragraph{Execution phase}
The winning \bidder execute \algorithm{SubmitWork} (\Cref{fig:overview3}-\ding{205}) and submits hashes of the encrypted files to store. They then contact the corresponding \worker off-chain, prove they are the rightful winner by showing their credential, and directly transmits the file replica $\mathcal{R}^\mathcal{D}_{ek}$ to the \worker. The \worker can then start generating proofs of useful work, submit them to the blockchain, and claim rewards from the winner's deposit. Verifying the proof of useful work and releasing the payment is an integral part of the underlying blockchain, and is out of the scope of this work. 

\section{Discussion}
\rname involves \worker specific contracts that offer resources in the form of a single item.
In practice more sophisticated auctions are of interest where multiple \workers offer their resources in the form of multiple items while \bidders express their bids for a subset of them on the same contract; such ambition comes at the challenges of \first auction execution, and \second bidding privacy and security. 

With respect to the auction execution, it has been shown that combinatorial auction can be modelled as the set packing problem, meaning that they are NP-hard and there is no polynomial-time algorithm for finding the optimal allocation. A solution would be to consider a system where \bidders are associated into up to a single item. That would lead to a computationally feasible and efficient assignment, i.e., multi-item unit demand auction~\cite{andersson2013multi} setting, on the expense however of \bidders' flexibility in expressing their preferences. On the other hand, the \bidders are expected to submit a vector of bids, \ie one bid for each offered item. Although the \algorithm{Deposit} function can be easily scaled up, \ie by putting as a deposit the sum of bids or the maximum bid in a vector, the analysis of such vectors of bids could reveal information on the bidder andcreate a privacy challenge that has to be addressed.

%% file: sections/security.tex
\section{Protocol Analysis} \label{sec:analysis}
We argue how \rname achieves the design goals described in \Cref{sec:goals}.
\vspace{-4pt}
\paragraph{Hidden Minimum Price} Upon the execution phase of the auction, \rname guarantees that a \worker will offer their resources at a price higher than the minimum price or not offer their resources at all.
\vspace{-4pt}
\paragraph{\Bidders privacy} \rname takes advantage of \coconut's unlinkability property to break the link between the deposit of coins, the commit of bids, and the withdraw of the coins. As a result, \bidders can submit bids on auctions without revealing their identity, yet proving possession of a valid credential. 
\vspace{-4pt}
\paragraph{Bids privacy} Bids are kept private until $t_{commit}$; the zero-knowledge property of \coconut credentials implies that no information about $v$ is revealed while committing to a bid (\Cref{fig:overview2}-\ding{202}).
\vspace{-4pt}
\paragraph{Bids binding} \Bidders are bound to their bids as they are first required to commit to their bid (\Cref{fig:overview2}-\ding{202}), and then to open the commitment (\Cref{fig:overview2}-\ding{203}). \Bidders cannot open the commitment to another value than the previously committed as this implies forging the \coconut credentials.
\vspace{-4pt}
\paragraph{Public auditability} \rname is implemented as a smart contract; its correct execution can be verified by any 3$^{rd}$ party by taking advantage of the public auditability of the underlying smart contract plateform.
\vspace{-4pt}
\paragraph{Fairness} No single \authority can create credentials and steal all the coins in the buffer account of the smart contract---the threshold property of \coconut implies that adversaries need to corrupt an arbitrarily large set of authorities for this attack to be possible. \Bidders cannot participate to the auction phase (\Cref{fig:overview2}) without paying a deposit to receive a valid credential; setting $t_{commit} < t_{reveal}$ forces the \bidders to commit to their bid before revealing it, preventing any 3$^{rd}$ party from seeing other \bidder's bid before committing to a value. \Bidders dropping out after  committing a bid (and never revealing it) are financially penalized as they cannot withdraw their coins. \coconut provides blind issuance which allows \bidders to obtain a credential on the sequence number $s$ without the \authorities learning its value. Without blindness, any authority seeing $s$ could potentially race the \bidders, and withdraw the coins of the credential---blindness prevents \authorities from stealing the coins. Keeping a spent list of all group elements $\zeta$ prevents double-spending attacks~\cite{karame2012double} without revealing the sequence number $s$; this prevents an attacker from exploiting a race condition upon withdrawing and which may lock \bidders coins.
\vspace{-4pt}
\paragraph{Non-interactivity} \Bidders do not interact with each other. During the preparation phase (\Cref{fig:overview1}), \Bidders only interact with a subset of the the \authorities to receive a credential; during the auction phase (\Cref{fig:overview2}), they only interact with the smart contract; during the execution phase (\Cref{fig:overview3}), \bidders only interact with the \workers or with the smart contract to withdraw their coins.
\vspace{-4pt}
\paragraph{Censorship resistance} The decentralized nature of the underlying smart contract platform makes the \rname smart contract resilient to censorship. Furthermore, a small subset of \authorities cannot block the issuance credentials---the service is guaranteed to be available as long as at least a threshold number of authorities are running.
\vspace{-4pt}
\paragraph{Distributed authority} \rname introduce no single trusted 3$^{rd}$ party; the \rname contract is executed on a decentralized smart contract platform, and \coconut allows threshold issuance of credentials.
\vspace{-4pt}
\paragraph{Auction's economic properties}
An auction satisfies all those properties only under the condition of \textit{price-taker participants}~\cite{myerson1983efficient}, \ie both \bidders and \workers have no impact on the auction prices. In the single item auctions we consider here, it has been proven that Vickrey auctions possesses all these desired attributes based on the assumption of ``sealed bids", where neither \bidders nor \workers have information about the state of the auction~\cite{vickrey1961counterspeculation}. \rname through its privacy properties provides a technical implementation of the ``sealed bids" assumption which prevents price manipulations.

%% file: sections/evaluation.tex
\section{Implementation \& Evaluation}

\begin{table}[t]
\centering
\begin{tabular}{L C C C}
    \toprule
    \multicolumn{4}{l}{\sysname \chainspace smart contract}\\
    \small\bf Operation & \small\bf \boldmath $\mu$ [ms] & \small\bf \boldmath$\sqrt{\sigma^2}$ [ms] & \bf size [kB]\\
    \midrule
    \textsf{Create} [g] & 28.433 & $\pm$ 0.214 & $\sim1.8$\\ 
    \textsf{Create} [c] & 0.0148 & $\pm$ 0.002 & -\\ 
    \textsf{Commit} [g] & 194.243 & $\pm$ 0.410 & $\sim2.7$\\ 
    \textsf{Commit} [c] & 355.852 & $\pm$ 15.880 & -\\
    \textsf{Reveal} [g] & 205.656 & $\pm$ 5.659 & $\sim2.7$\\ 
    \textsf{Reveal} [c] & 351.192 & $\pm$ 8.514 & -\\
    \textsf{Withdraw} [g] & 188.925 & $\pm$ 2.084 & $\sim2.6$\\ 
    \textsf{Withdraw} [c] & 336.533 & $\pm$ 4.490 & -\\
    \textsf{SubmitWork} [g] & 197.399 & $\pm$ 6.537 & $\sim2.7$\\ 
    \textsf{SubmitWork} [c] & 368.948 & $\pm$ 13.116 & -\\
    \bottomrule
\end{tabular}
\caption{\footnotesize Timing and transaction size of the \sysname \chainspace smart contract (described in \Cref{sec:smart-contract}), measured over 10,000 runs. The transactions are independent of the number of authorities. The notation [g] denotes the execution the procedure and [c] denotes the execution of the checker. The \textsf{Deposit} function is not implemented as it is identical to the tumbler application described in \coconut~\cite{coconut}.}
\label{tab:evaluation}
\end{table}
We provide an open-source implementation\footnote{\url{https://github.com/asonnino/coconut-chainspace}} of the \sysname smart contract presented in \Cref{sec:smart-contract} for \chainspace~\cite{chainspace}. Our implementation does not enforce conditions on timers $t_{commit}$ and $t_{reveal}$ as \chainspace currently does not provide functions to check block heights. We write our prototype in about 450 lines of Python code using an open source Python implementation of Coconut\footnote{\url{https://github.com/asonnino/coconut}}, which reply on petlib\footnote{\url{https://github.com/gdanezis/petlib}} and pblib\footnote{\url{https://github.com/gdanezis/bplib}}; the bilinear pairing is
defined over the Barreto-Naehrig curve\footnote{\url{https://tools.ietf.org/id/draft-kasamatsu-bncurves-01.html}}, using OpenSSL as arithmetic backend. \Cref{tab:evaluation} provides the timing and transaction size for each function of the smart contract; each experiment is the result of 100 runs measured on a commodity laptop (a MacBook Pro 13' 2.7 GHz Intel Core i7, running macOS Mojave). As expected, both the procedure and the checker of \textsf{Create} are extremely fast as they don't involve cryptographic operations. The checker of \textsf{Commit}, \textsf{Reveal}, \textsf{Withdraw}, and \textsf{SubmitWork} take on average the same (and the longest) time as their core operation is to verify credentials validity; verifying credentials takes the longest time due to pairing operations~\cite{coconut}.  The \textsf{Deposit} function is not implemented as it is identical to the tumbler application described in \coconut~\cite{coconut}.

%% file: sections/related.tex
\section{Related Work}\label{sec:related}

\begin{table*}[t]
\centering
\begin{tabular}{  L  C  C  C  C C C}
\toprule
\bf System & \bf Bids Privacy & \bf Bidders Privacy & \bf Bidders Non-Interactivity & \bf Distributed Authority & \bf Trusted Hardware & \bf Public Auditability\\ 
\midrule    

ShadowEth~\cite{yuan2018shadoweth} & \cmark & \xmark  & \cmark & \H & Intel SGX~\cite{costan2016intel} & \cmark\\
Hawk~\cite{kosba2016hawk} & \cmark & \xmark&  \cmark & \M & None  & \cmark\\
Strain~\cite{blass2018strain} & \cmark & \xmark & \xmark & \L & None  & \cmark\\
Galal~\etal~\cite{galal2018succinctly} & \cmark & \xmark & \xmark & \L & None  & \cmark\\
Bogetoft~\etal~\cite{bogetoft2009secure} & \cmark & \xmark & \cmark & \H & None  & \xmark\\
Filecoin~\cite{filecoin} & \xmark & \xmark & \xmark & \H & None & \cmark\\
\rowcolor{verylightgray}
\textbf{\rname} & \cmark & \cmark & \cmark & \H & None & \cmark \\

\bottomrule
\end{tabular}
\caption{\footnotesize Comparison security properties achieved by different systems discussed in this section. The decentralization property reads as follows; \L : relies on a trusted 3$^{rd}$ party, \M : relies on a trusted 3$^{rd}$ party for only one (or some) of the properties described in \Cref{sec:goals}, \H : does not rely on any trusted 3$^{rd}$ party.}
\label{tab:related}
\vspace{-15pt}
\end{table*}

There are several frameworks that target hiding private data submitted to a public ledger. Hawk \cite{kosba2016hawk}  divides Smart Contract into public and private parts and secure private input using zero-knowledge proofs, but requires a centralized trusted manager to operate. Furthermore, ShadowEth \cite{yuan2018shadoweth} allows processing confidential Smart Contract data using Trusted Execution Environments (TEE). However, such a scheme requires users to trust the hardware vendor and can expose the system to TEE's vulnerabilities \cite{sgx_bad1, sgx_bad2}. 
On the subject of sealed-bid auctions, Blass and Kerschbaum \cite{blass2018strain} proposed Strain, that preserve bids privacy against malicious participants. Strain uses a two-party comparison protocol, but has a flaw that reveals the order of bids. Furthermore, running protocols involving MPC on blockchain is not efficient due to extensive computations and the number of rounds involved.
Furthermore, Galal and Youssef \cite{galal2018verifiable} presented a protocol that ensures public verifiability, privacy of bids, and fairness. However, the solution scales badly with the number of bidders and relies on random number retrieved from blockchains that are not proven to be secure. This scheme was improved in \cite{galal2018succinctly} using zk-SNARKS, but still relies on a centralized party for zero-knowledge proofs and does not protect bidders identity. 
An alternative approach was proposed by Bogetoft \etal~\cite{bogetoft2009secure}. The system uses a multiparty computation to perform auctions on encrypted bids. However, such a scheme reveals final assignment between bidders and object and lacks transparency. 
Currently, Filecoin \cite{filecoin} does not implement an automated system assigning clients to storage nodes. Users are required to chose storage nodes manually and offers are publicly posted on the blockchain. Other industrial system such Golem \cite{golem}, iExec \cite{iexec} or SONM \cite{sonm} either do not specify their requester---worker assignment technique rely on similar, insecure solutions. All those platform could use \rname to increase their level of security and automatically determine optimal price for services. 
We summarize discussed solutions and their security features in \Cref{tab:related}.

%% file: sections/limitations.tex
\section{Limitations} \label{sec:limitations}
\rname inherits several limitations of \coconut which acts as the underlying credential scheme. These limitations are beyond the scope of this work, and deferred to future work. \rname is vulnerable if more than the threshold number of \authorities are malicious; colluding \authorities could create credentials to steal all the coins in the buffer of the smart contract. Note that \bidders’ privacy is still guaranteed under colluding \authorities, or an eventual compromise of their keys.

The smart contract implementation of \sysname described in \Cref{sec:smart-contract} does not scale to a large number of users as as each commitment and bid is kept on-chain. A potential solution is to defer the auction logic to state channel~\cite{miller2017sprites}, and only seal the final result of the auction on the blockchain. Moreover, \rname inherits from any scalability limitation of the underlying smart contract platform.


The winner of the auction may invoke \algorithm{SubmitWork}, but refuse to transfer data to the worker. This prevents the worker from claiming the reward and leave their resources unused until the next auction. This issue is inherited from Filecoin and can be mitigated with additional mechanisms assuring fair exchange of digital goods~\cite{dziembowski2018fairswap}. 


%% file: sections/conclusion.tex
\section{Conclusion and Future Work}
We presented \rname --- a system for determining optimal prices for services in a shared economy environment and automatic assignments of requesters to the most optimal workers. \rname allows to securely perform sealed-bid auction on a blockchain using anonymous credentials and zero-knowledge proofs of knowledge. Our system allows workers to specify a minimal price for their services, protects users bids as well as the identity of the bidders. Contrary to the previous work, \rname does not rely on a trusted 3$^{rd}$ party to issue credentials, but rather on a set of entities that can be freely chosen by users. The distributed authorities issue only partial credentials that are merged locally by each users protecting the system from a subset of malicious authorities. We showed how \rname can be deployed in the Filecoin network and adapted to other shared economy systems operating on blockchain. As a part of future work, we plan to extend our system to securely support requesters' evaluation of multiple items and protect against data analysis attacks. Furthermore, we will investigate scalability of our solution with large networks and better integration with additional platforms.

%% file: sections/ack.tex
\section*{Acknowledgements} Alberto Sonnino is supported by the EU H2020 DECODE project under grant agreement number 732546 as well as \texttt{chainspace.io}. Michał Król and Argyrios G. Tasiopoulos are supported by the EC H2020 ICN2020 project under grant agreement number 723014. Ioannis Psaras is supported by the EPSRC INSP Early Career Fellowship under grant agreement number EP/M003787/1.